\documentclass[a4paper,11pt]{article}
\usepackage{pos}

\title{The Surface Array planned for IceCube-Gen2}
 \ShortTitle{IceCube-Gen2 Surface Array}

\author{The IceCube-Gen2 Collaboration \\{\normalsize \normalfont(a complete list of authors can be found at the end of the proceedings)}}

\emailAdd{frank.schroeder@icecube.wisc.edu}

\abstract{IceCube-Gen2, the extension of the IceCube Neutrino Observatory, will feature three main components: an optical array in the deep ice, a large-scale radio array in the shallow ice and firn, and a surface detector above the optical array. Thus, IceCube-Gen2 will not only be an excellent detector for PeV neutrinos, but also constitutes a unique setup for the measurement of cosmic-ray air showers, where the electromagnetic component and low-energy muons are measured at the surface and high-energy muons are measured in the ice. As for ongoing enhancement of IceCube’s current surface array, IceTop, we foresee a combination of elevated scintillation and radio detectors for the Gen2 surface array, aiming at high measurement accuracy for air showers. The science goals are manifold: The in-situ measurement of the cosmic-ray flux and mass composition, as well as more thorough tests of hadronic interaction models, will improve the understanding of muons and atmospheric neutrinos detected in the ice, in particular, regarding prompt muons. Moreover, the surface array provides a cosmic-ray veto for the in-ice detector and contributes to the calibration of the optical and radio arrays. Last but not least, the surface array will make major contributions to cosmic-ray science in the energy range of the transition from Galactic to extragalactic sources. The increased sensitivities for photons and for cosmic-ray anisotropies at multi-PeV energies provide a chance to solve the puzzle of the origin of the most energetic Galactic cosmic rays and will serve IceCube’s multimessenger mission.

\vspace{4mm}
{\bfseries Corresponding authors:}
Frank G.~Schroeder$^{1,2}$\\
{$^{1}$ \itshape Bartol Research Institute, Department of Physics and Astronomy, University of Delaware, Newark DE, USA}\\
{$^{2}$ \itshape Institute for Astroparticle Physics (IAP), Karlsruhe Institute of Technology (KIT), Karlsruhe, Germany}\\[4mm]
$^*$ Presenter

\FullConference{37$^{\rm{th}}$ International Cosmic Ray Conference (ICRC 2021)\\
		July 12th -- 23rd, 2021\\
		Online -- Berlin, Germany}

}

\begin{document}
\maketitle

\section{Introduction}
Building on the history of the IceTop surface array above the deep neutrino detector of IceCube, we want to expand this successful approach to IceCube-Gen2 \cite{Aartsen:2020fgd}.
IceCube-Gen2, the next generation of the IceCube Neutrino Observatory at the South Pole, will consist of three detector arrays: a deep optical array consisting of strings with optical modules in the ice which will detect atmospheric muons of TeV to PeV energies as well as different types of neutrino-induced signals, a radio array consisting of antennas in the shallow ice and firn to extend neutrino measurements to the EeV energy range, and a surface array comprised of one hybrid surface station above each optical string.

The purpose of the surface array is to support IceCube's multi-messenger mission of searching for the sources of ultra-high-energy cosmic rays. 
These are predominantly of extragalactic origin, but depending on the scenario for the transition from Galactic to extragalactic origin of cosmic rays, some of the cosmic rays observed with EeV energies may still be of Galactic origin \cite{Aab:2020rhr, Aab:2020xgf, Mollerach:2017idb}. 
While neutrino detection promises to reveal extragalactic EeV sources, the higher accuracy and exposure of the Gen2 surface array compared to previous arrays will contribute to a better understanding of these most energetic Galactic cosmic rays, which is essential for a complete and consistent picture of ultra-high-energy particle astrophysics \cite{Schroeder:2019TB, Astro2020_GCR_WhitePaper}. 

The IceCube-Gen2 surface array will also extend those capabilities of IceTop directly important for the neutrino measurements.
It will provide a veto for downgoing events, improve the understanding of hadronic interactions in air showers, in particular, regarding the flux of prompt muons, and it can be used to cross-check the calibrations of the optical and radio arrays.

\section{Detector Design}
\begin{figure}[t]
    \centering
    \includegraphics[width=0.59\linewidth]{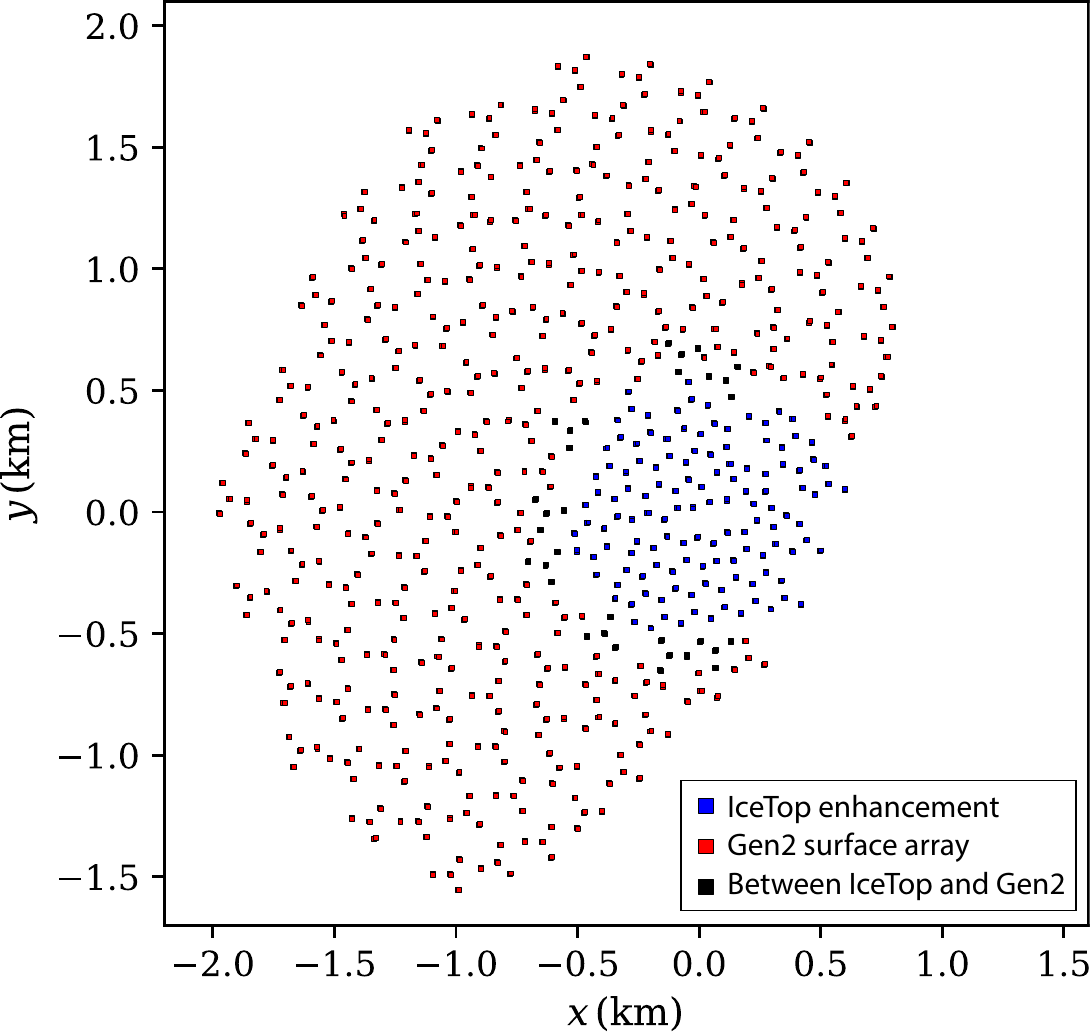}
    \hfill
    \includegraphics[width=0.39\linewidth]{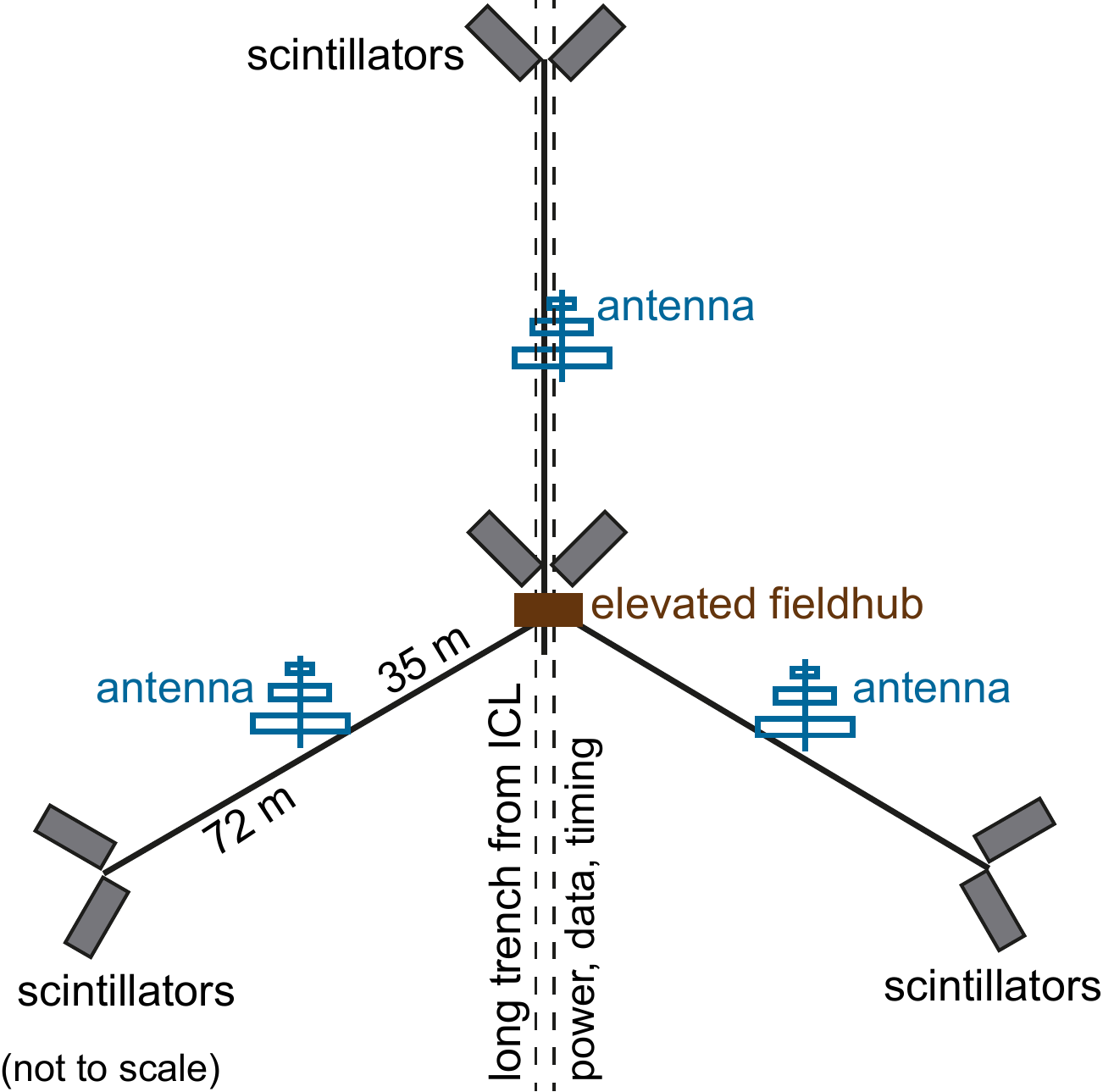}
    \caption{Layout of the IceCube-Gen2 Surface Array (left) and station design (right). The layout builds on the IceTop enhancement using the same station design of 4 pairs of scintillators and 3 surface radio antennas connected to a local DAQ in an elevated fieldhub. There will be one station on top of each string of the Gen2 optical array in addition to a few stations filling gaps to the planned surface enhancement of IceTop.
    }
    \label{fig:layout}
\end{figure}

\begin{figure}[t]
    \centering
    \includegraphics[width=0.68\linewidth]{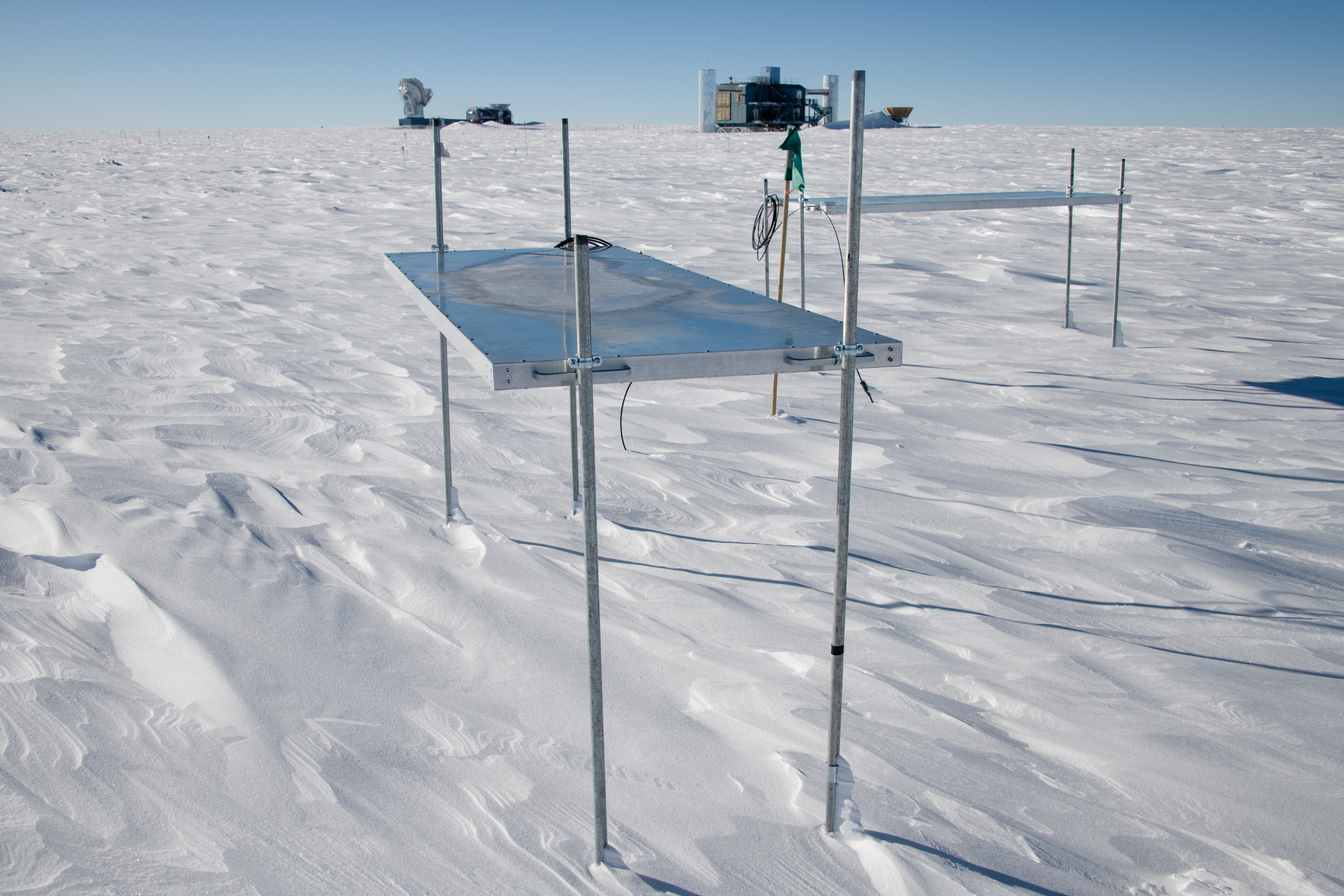}
    \hfill
    \includegraphics[width=0.303\linewidth]{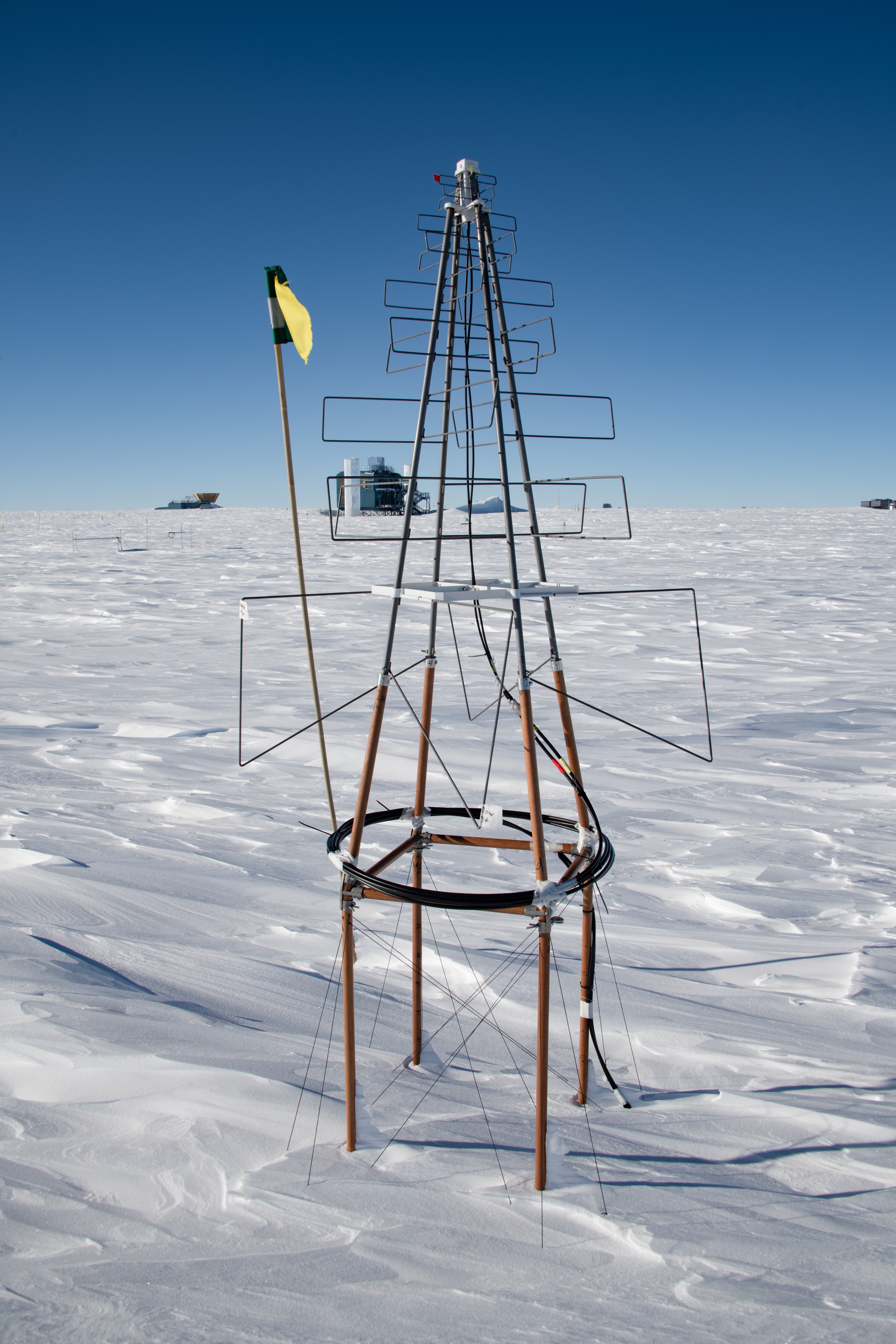}
    \caption{Surface detectors of the prototype station deployed in January 2020 after the winter; left: one of the four pairs of scintillation panels; right: one of the three radio antennas. No build-up of snow is observed except of the typical snow accumulation of approx.~$20\,$cm per year. The mount structures are designed such that the panels and antennas can be raised every few years.
    }
    \label{fig:detectors}
\end{figure}

The layout of the IceCube-Gen2 surface array extends the planned enhancement of IceTop \cite{Haungs:2019ylq} using the same station design (Fig.~\ref{fig:layout}). 
Each station consists of eight scintillation panels arranged in pairs and three radio antennas placed half way along the trenches to the scintillation panels. 
The main purpose of the scintillators is to provide a low detection threshold of about $0.5\,$PeV, and the main purpose of the radio antennas is to increase the accuracy for the mass composition at the energy range above $100\,$PeV.
Based on a history of several prototypes, a complete prototype station was installed at the South Pole in 2020 and is successfully measuring air showers with all detectors (see \cite{SurfaceArray_ICRC2021} for details). 
The detectors are elevated to avoid snow coverage (Fig.~\ref{fig:detectors}) because snow otherwise would increase the detection threshold and cause systematic uncertainties in the interpretation of the measurements.
Due to the positive experience with the prototype station, we plan to use the same detectors design with just smaller improvements for the Gen2 surface array.

The scintillation panels have a detection area of $1.5$m$^2$ and consist of plastic scintillation stripes that are coupled to optical fiber connected to a SiPM \cite{IceScint_ICRC2019}. 
A local electronics 'microDAQ' digitizes the signal in three amplification channels and allows for a regular calibration on the signal strengths of a minimum ionizing particle (MIP). 
In these units of MIP, the panels will cover a dynamic range of more than three orders of magnitude starting with a detection threshold of about $0.5$\,MIP.

The surface antennas will be of type SKALA v2 \cite{SKALAv2}, which provide a smooth sky coverage over the entire frequency band of interest, which is about $70-350\,$MHz. 
Each antenna features two polarization channels equipped with a low-noise amplifier that ensures a high measurement precision contributing only about $40\,$K thermal noise. 
Thus, over the entire measurement band the unavoidable Galactic noise will be the relevant background and, at the same time, also a calibration source.
Unlike at most other locations on Earth, human-made RFI at the South Pole is very low which enables to build a radio detector including the FM band.

The data of the eight scintillators of one station are collected by a local DAQ inside an elevated fieldhub in the center of each station. 
This will be a more advanced version of the TAXI system currently used for the IceTop enhancement \cite{TAXI_ICRC2021}.
The same DAQ digitizes the radio wavefronts and stores them when receiving a trigger, e.g., by the scintillators. 
Data will then be sent to the central DAQ at the IceCube Lab (ICL) where they are merged with other data streams.

Finally, the addition of IceAct telescopes \cite{IceAct_ICRC2021} is planned at four of the stations to increase the measurement accuracy for cosmic rays of lower energy. 

Together with the IceTop enhancement, the IceCube-Gen2 surface array will cover an area of about $6\,$km$^2$ with a total of 162 stations: 1 on top of each of the strings of the optical array, 32 of the IceTop enhancement, and 8 stations to avoid a higher threshold for events falling into the small gap that would otherwise exist between the two surface arrays.  
While the geometric surface area increases 'only' by a factor of 8, the geometric aperture for air-shower events which also feature an in-ice signal increases by more than a factor of 30 compared to IceCube.
This is because of the largely increased range of zenith angles of possible coincidences. 
In particular these events exploit a unique feature provided by IceCube-Gen2: the simultaneous measurement of MeV electromagnetic particles and GeV muons at the surface and of TeV muons in the ice. 
For high-energy events, the radio antennas provide in addition a measurement of the position of the shower maximum and of the calorimetric energy (they also do at the IceTop enhancement, but due to the limited zenith coverage only for a tiny fraction of the events coincident with the in-ice detector).
Overall, this unique combination of shower observables including the deep detector will make IceCube-Gen2 the most accurate detector for high-energy Galactic cosmic rays.

\begin{table}[p]
    \caption{Overview of the science case of the IceCube-Gen2 surface array.}
    \centering
    \begin{tabular}{p{0.9\linewidth}}
        \textbf{Veto for the IceCube-Gen2 Optical Array}\\
        \hline
        \vspace{-0.2cm} $\bullet$\,\,\,Increased sensitivity for downgoing neutrinos for a large range of zenith angles\\
        \vspace{-0.2cm} $\bullet$\,\,\,Check high-energy neutrino candidates identified by the real-time alert system\\
        \vspace{-0.2cm} $\bullet$\,\,\,Test applicability of surface antennas as veto for $>10\,$PeV inclined neutrino candidates\\
        \\
        \textbf{Hadronic Interactions in Air Showers}\\
        \hline
        \vspace{-0.2cm} $\bullet$\,\,\,Investigate transition from conventional to prompt muon fluxes around $0.5 - 1\,$PeV\\
        \vspace{-0.2cm} $\bullet$\,\,\,Scrutinize interaction models by muon spectroscopy (GeV at surface, TeV in ice)\\
        \vspace{-0.2cm} $\bullet$\,\,\,Extend muon measurements at surface to $0.5$\,EeV for overlap with AMIGA at Auger\\
        \\
        \textbf{Most energetic Galactic Cosmic Rays (and transition to extragalactic CRs)}\\
        \hline
        \vspace{-0.2cm} $\bullet$\,\,\,Unprecedented accuracy for primary mass by combination of surface and deep detectors\\
        \vspace{-0.2cm} $\bullet$\,\,\,Extend energy range of large-scale dipole anisotropy at high statistical significance\\
        \vspace{-0.2cm} $\bullet$\,\,\,Increase IceCube's exposure for PeV photon searches by an order of magnitude\\
        \\
        \textbf{Cross-check Calibration of in-ice Detectors (optical and radio arrays)}\\
        \hline
        \vspace{-0.2cm} $\bullet$\,\,\,In-situ measurement of cosmic-ray flux and muon tagging for in-ice optical array\\
        \vspace{-0.2cm} $\bullet$\,\,\,Provide energy estimate of vertical showers for calibration of in-ice radio array\\
        \vspace{-0.2cm} $\bullet$\,\,\,Cross-calibrate absolute energy scale for cosmic-ray air showers by radio antennas\\
        
    \end{tabular}
    \vspace{-0.1cm}
    \label{tab:sciencecase}
\end{table}

\begin{figure}[p]
    \centering
    \includegraphics[width=0.69\linewidth]{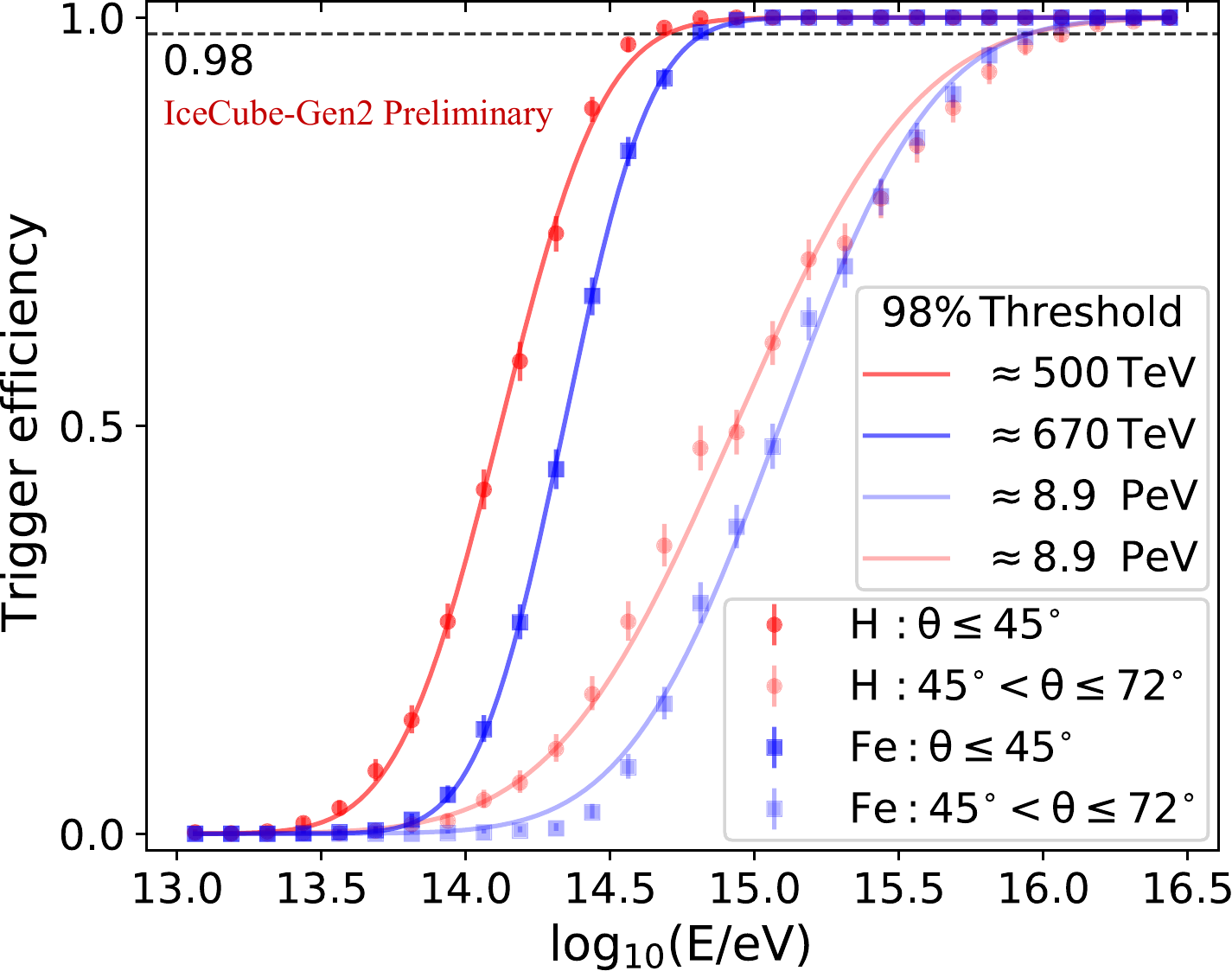}
    \caption{Simulated threshold of the scintillators for proton and iron primaries assuming a trigger threshold for the individual panels of $0.5\,$MIP and a fivefold coincidence over the array.
    }
    \label{fig:scintThreshold}
\end{figure}

\begin{figure}[t]
    \centering
    \includegraphics[width=0.49\linewidth]{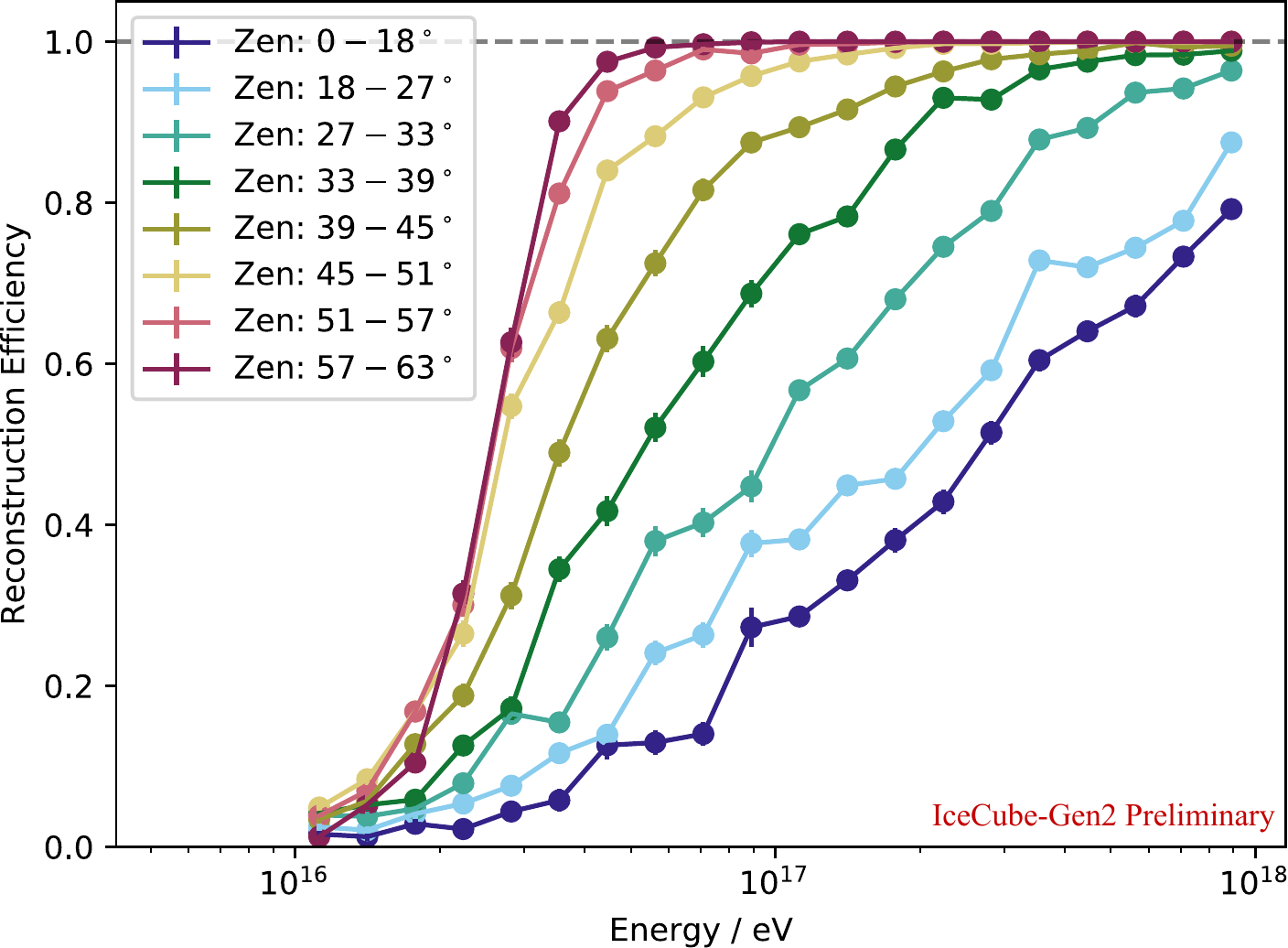}
    \hfill  
    \includegraphics[width=0.49\linewidth]{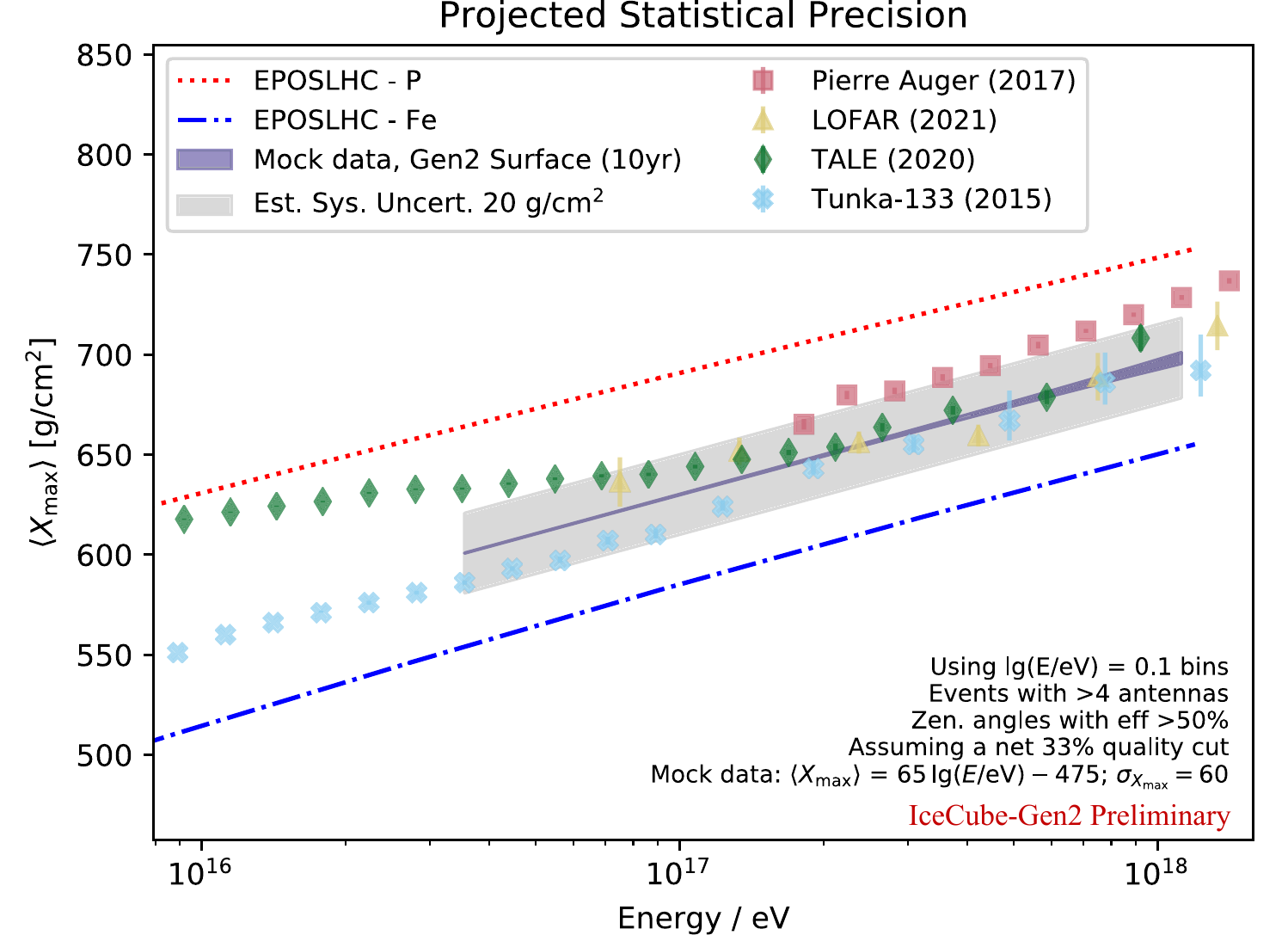}
    \caption{Left: Threshold of the surface radio antennas; assuming Galactic noise as background, the signal-to-noise ratio in the individual antennas needs to exceed the value that corresponds to a $98\,\%$ rejection of background waveforms in at least three antennas.
    Right: Statistical precision expected for radio measurements of the mean $X_\mathrm{max}$ for 10 years of mock data assuming that the statistical fluctuations around the mean are of $60\,$g/cm$^2$ and that  one third of the events with at least five antennas can be used for $X_\mathrm{max}$ measurements (simulations beyond $1\,$EeV are not yet completed).
    }
    \label{fig:radioThreshold}
\end{figure}

\section{Science Case}
The IceCube-Gen2 surface array supports and enriches the multi-messenger mission of IceCube-Gen2 with a broad science case (see Tab.~\ref{tab:sciencecase}). 
These science cases fall into three overlapping categories: supporting the neutrino detection directly, such as the anti-coincidence veto and the cross-checks of the in-ice calibration; supporting the neutrino detection indirectly by a better understanding of atmospheric lepton fluxes in the ice and, in particular, of the prompt muon fluxes; supporting the search for the cosmic-ray sources by other messengers, in particular by a more accurate measurement of the most energetic Galactic cosmic rays and by the search for PeV photons. 

For the \textbf{veto}, the surface array will increase the sensitivity and purity for down-going neutrino candidates. 
The fill factor of active detection area per geometric area will be about two third of what IceTop provided. 
However, due to the avoidance of snow coverage, the veto threshold is still expected to be similar to IceTop \cite{Gen2surfaceSims_ICRC2021}. On the one hand, the veto will be used in the same direct way as with IceTop for IceCube, but for a larger range of zenith angles \cite{Aartsen:2016oji}. 
On the other hand, as also done with IceTop already, we will continue to check potential high-energy real-time alters for surface signals to obtain a purer sample of neutrino candidates. 
Finally, we plan to test the option to what extend the surface antennas can be used to veto parent showers of $> 10\,$PeV very inclined down-going neutrino candidates.

The combination of a surface array with a deep detectors enables some unique ways to test \textbf{hadronic interaction models} for a better understanding of the forward particle physics in high-energy cascades.
A long standing problem is the muonic component of air showers. 
All available hadronic interaction models predict to few muons at higher energies \cite{Cazon:2019ud}. 
In addition, there are huge uncertainties regarding the flux of prompt leptons (muons and neutrinos produced in charm decays).
IceCube measures a few PeV muons per year in the ice, and a better understanding of their origin is crucial for their interpretation.

Depending on the scenario, a transition from predominantly conventional to prompt muons is expected between approximately $0.5$ to $1\,$PeV \cite{Aartsen:2015nss}.
While IceCube has measured several muons in this energy range, due to the small zenith range of surface coincidences, it is not possible to obtain information of the parent showers for a significant number of these events. 
This will change entirely with the more than 30 times larger aperture for such events in IceCube-Gen2 and requires a low enough threshold of the surface array. 
According to Ref.~\cite{Fedynitch:2018cbl}, a PeV muon can carry a significant fraction of the primary particle if it is a proton. 
Therefore, to enable a study of the transition energy from conventional to prompt muons, the surface array is designed such that it enables a threshold of $0.5\,$PeV for proton induced showers (Fig.~\ref{fig:scintThreshold}).

Muons can also be measured at the surface. While generally the surface signal is dominated by electromagnetic particles, muons become more prominent with increasing distance to the shower axis. 
This is already used to measure the muon density in $600\,$m and $800\,$m distance with IceTop \cite{Gonzalez:2019epd}. 
By statistics alone these measurements can be extended in energy to several $100\,$PeV which will close the gap to and provide overlap with AMIGA measurements \cite{Aab:2020frk}.
Moreover, due to the larger zenith range of in-ice coincidences, it will be possible to perform combined analyses of the TeV muons in the ice and GeV surface muons.
Such a muon spectroscopy provides additional ways to test hadronic interactions. 
Finally, the inclusion of the calorimetric radio information on a per-event basis will enable to test hadronic interaction models for individual showers instead of just using statistical averages over large event samples.

Regarding \textbf{cosmic-ray physics}, the combination of a surface particle, surface radio, and in-ice detector will provide unprecedented accuracy for the mass composition, in particular at high energies where the radio detection can provide a measurement of the atmospheric depth of the shower maximum, $X_\mathrm{max}$ (Fig.~\ref{fig:radioThreshold}) \cite{HuegeReview2016, SchroederReview2016}.
Radio arrays have already demonstrated to provide a competitive  $X_\mathrm{max}$ resolution \cite{TunkaRexPRD2018, LOFARNature2016}, and the combination with the additional muon information of the surface and deep detectors promises unprecedented resolution for the estimation of the primary mass \cite{Holt:2019fnj}.
This is important because different scenarios for the Galactic-to-extragalactic transition can be distinguished by the mass composition. 
Moreover, the higher aperture of the Gen2 surface array will enable to extend IceTop's anisotropy measurements to higher energies and increases the statistical sensitivity to PeV photons. 
At the same time, the larger aperture for events coincident with the in-ice detector in combination with the scintillator array will allow for the development of better gamma-hadron separation methods \cite{Aartsen:2019sid}. 
As the first source of PeV photons has been discovered recently by LHAASO \cite{LHAASO_Nature2021}, though outside the field of view of IceCube-Gen2, it may be just a question of observation time until PeV photons are also identified by IceCube-Gen2.

For completeness, the surface array will also be used to cross-check the calibration for optical and radio arrays, e.g., by providing accurate measurements of air showers that are also observed by their radio or optical signals in the ice. 
Finally, radio measurements of air-showers have been proven a useful tool to cross-calibrate the absolute energy scale over different experiments \cite{Apel:2016gws}.

\section{Conclusion}
The design of the IceCube-Gen2 surface array builds on the plans for the enhancement of the IceTop array and the positive experience with the corresponding prototype station at the South Pole.
Featuring one hybrid station of elevated scintillation and radio detectors at each optical string, the surface array will feature a threshold around $0.5\,$PeV and provide air-showers measurements over at least four orders of magnitude in energy. 
By the low energy threshold, the surface array will provide an effective veto for high-energy neutrino candidates. In addition, the low threshold of the surface array in combination with the in-ice detector will enable a better understanding of transition from dominantly conventional to prompt muons around $1\,$PeV by analysing coincident events.
Practically all cosmic-ray studies will benefit from the larger exposures compared to IceTop and IceCube.
However, many open questions regarding the most energetic Galactic cosmic rays require not just more statistics, but an increase in the accuracy of the mass of the primary particles \cite{Astro2020_GCR_WhitePaper}.
This increase in per-event accuracy will be provided by a world-unique combination of a hybrid surface array of particle and radio detectors with a deep optical detector.

\bibliographystyle{ICRC}
\bibliography{references}

\providecommand{\href}[2]{#2}\begingroup\raggedright\begin{thebibliography}{10}

\bibitem{Aartsen:2020fgd}
{\bfseries IceCube-Gen2} Collaboration, M.~G. Aartsen {\em et~al.}
  \href{http://dx.doi.org/10.1088/1361-6471/abbd48}{{\em J. Phys. G} {\bfseries
  48} no.~6, (2021) 060501}.

\bibitem{Aab:2020rhr}
{\bfseries Pierre Auger} Collaboration, A.~Aab {\em et~al.}
  \href{http://dx.doi.org/10.1103/PhysRevLett.125.121106}{{\em Phys. Rev.
  Lett.} {\bfseries 125} no.~12, (2020) 121106}.

\bibitem{Aab:2020xgf}
{\bfseries Pierre Auger} Collaboration, A.~Aab {\em et~al.}
  \href{http://dx.doi.org/10.3847/1538-4357/ab7236}{{\em Astrophys. J.}
  {\bfseries 891} (2020) 142}.

\bibitem{Mollerach:2017idb}
S.~Mollerach and E.~Roulet
  \href{http://dx.doi.org/10.1016/j.ppnp.2017.10.002}{{\em Prog. Part. Nucl.
  Phys.} {\bfseries 98} (2018) 85--118}.

\bibitem{Schroeder:2019TB}
F.~Schroeder \href{http://dx.doi.org/10.22323/1.358.0030}{{\em PoS} {\bfseries
  ICRC2019} (2019) 030}.

\bibitem{Astro2020_GCR_WhitePaper}
F.~G. Schroeder {\em et~al.} {\em BAAS} {\bfseries 51} (2019) 131. Astro2020
  Science White Paper.

\bibitem{Haungs:2019ylq}
{\bfseries IceCube} Collaboration, A.~Haungs
  \href{http://dx.doi.org/10.1051/epjconf/201921006009}{{\em EPJ Web Conf.}
  {\bfseries 210} (2019) 06009}.

\bibitem{SurfaceArray_ICRC2021}
{\bfseries IceCube} Collaboration
  \href{http://dx.doi.org/10.22323/1.395.0314}{{\em PoS} {\bfseries ICRC2021}
  (2021) 314}. (these proceedings).

\bibitem{IceScint_ICRC2019}
{\bfseries IceCube} Collaboration, M.~Kauer {\em et~al.}
  \href{http://dx.doi.org/10.22323/1.358.0309}{{\em PoS} {\bfseries ICRC2019}
  (2019) 309}.

\bibitem{SKALAv2}
E.~de~Lera~Acedo {\em et~al.}
  \href{http://dx.doi.org/10.1109/ICEAA.2015.7297231}{{\em 2015 International
  Conference on Electromagnetics in Advanced Applications (ICEAA)} (09, 2015)
  839--843}.

\bibitem{TAXI_ICRC2021}
{\bfseries IceCube} Collaboration
  \href{http://dx.doi.org/10.22323/1.395.0225}{{\em PoS} {\bfseries ICRC2021}
  (2021) 225}. (these proceedings).

\bibitem{IceAct_ICRC2021}
{\bfseries IceCube} Collaboration
  \href{http://dx.doi.org/10.22323/1.395.0276}{{\em PoS} {\bfseries ICRC2021}
  (2021) 276}. (these proceedings).

\bibitem{Gen2surfaceSims_ICRC2021}
{\bfseries IceCube-Gen2} Collaboration
  \href{http://dx.doi.org/10.22323/1.395.0411}{{\em PoS} {\bfseries ICRC2021}
  (2021) 411}. (these proceedings).

\bibitem{Aartsen:2016oji}
{\bfseries IceCube} Collaboration, M.~G. Aartsen {\em et~al.}
  \href{http://dx.doi.org/10.3847/1538-4357/835/2/151}{{\em Astrophys. J.}
  {\bfseries 835} no.~2, (2017) 151}.

\bibitem{Cazon:2019ud}
L.~Cazon \href{http://dx.doi.org/10.22323/1.358.0214}{{\em PoS} {\bfseries
  ICRC2019} (2019) 214}.

\bibitem{Aartsen:2015nss}
{\bfseries IceCube} Collaboration, M.~G. Aartsen {\em et~al.}
  \href{http://dx.doi.org/10.1016/j.astropartphys.2016.01.006}{{\em Astropart.
  Phys.} {\bfseries 78} (2016) 1--27}.

\bibitem{Fedynitch:2018cbl}
A.~Fedynitch {\em et~al.}
  \href{http://dx.doi.org/10.1103/PhysRevD.100.103018}{{\em Phys. Rev. D}
  {\bfseries 100} no.~10, (2019) 103018}.

\bibitem{Gonzalez:2019epd}
{\bfseries IceCube} Collaboration, J.~G. Gonzalez
  \href{http://dx.doi.org/10.1051/epjconf/201920803003}{{\em EPJ Web Conf.}
  {\bfseries 208} (2019) 03003}.

\bibitem{Aab:2020frk}
{\bfseries Pierre Auger} Collaboration, A.~Aab {\em et~al.}
  \href{http://dx.doi.org/10.1140/epjc/s10052-020-8055-y}{{\em Eur. Phys. J. C}
  {\bfseries 80} no.~8, (2020) 751}.

\bibitem{HuegeReview2016}
T.~Huege \href{http://dx.doi.org/10.1016/j.physrep.2016.02.001}{{\em Phys.
  Rept.} {\bfseries 620} (2016) 1--52}.

\bibitem{SchroederReview2016}
F.~G. Schr\"oder \href{http://dx.doi.org/10.1016/j.ppnp.2016.12.002}{{\em Prog.
  Part. Nucl. Phys.} {\bfseries 93} (2017) 1--68}.

\bibitem{TunkaRexPRD2018}
{\bfseries Tunka-Rex} Collaboration, P.~A. Bezyazeekov {\em et~al.}
  \href{http://dx.doi.org/10.1103/PhysRevD.97.122004}{{\em Phys. Rev.}
  {\bfseries D97} no.~12, (2018) 122004}.

\bibitem{LOFARNature2016}
{\bfseries LOFAR} Collaboration, S.~Buitink {\em et~al.}
  \href{http://dx.doi.org/10.1038/nature16976}{{\em Nature} {\bfseries 531}
  (2016) 70}.

\bibitem{Holt:2019fnj}
E.~M. Holt, F.~G. Schr{\"o}der, and A.~Haungs
  \href{http://dx.doi.org/10.1140/epjc/s10052-019-6859-4}{{\em Eur. Phys. J.}
  {\bfseries C79} no.~5, (2019) 371}.

\bibitem{Aartsen:2019sid}
{\bfseries IceCube} Collaboration, M.~G. Aartsen {\em et~al.}
  \href{http://dx.doi.org/10.3847/1538-4357/ab6d67}{{\em Astrophys. J.}
  {\bfseries 891} (8, 2019) 9}.

\bibitem{LHAASO_Nature2021}
{\bfseries LHAASO} Collaboration, Z.~Cao {\em et~al.}
  \href{http://dx.doi.org/10.1038/s41586-021-03498-z}{{\em Nature} {\bfseries
  594} (2021) 33}.

\bibitem{Apel:2016gws}
{\bfseries Tunka-Rex, LOPES} Collaboration, W.~D. Apel {\em et~al.}
  \href{http://dx.doi.org/10.1016/j.physletb.2016.10.031}{{\em Phys. Lett.}
  {\bfseries B763} (2016) 179--185}.

\end{thebibliography}\endgroup
\vspace{5mm}
\noindent
\footnotesize{The preparations for IceCube-Gen2 are support by several agencies. Specific acknowledgement for this proceeding: This project has received funding from the European Research Council (ERC) under the European Union's Horizon 2020 research and innovation programme (grant agreement No 802729). Parts of the presented work have been supported by the U.S. National Science Foundation-EPSCoR (RII Track-2 FEC, award ID 2019597).}



\clearpage
\section*{Full Author List: IceCube-Gen2 Collaboration}

\scriptsize
\noindent
R. Abbasi$^{17}$,
M. Ackermann$^{71}$,
J. Adams$^{22}$,
J. A. Aguilar$^{12}$,
M. Ahlers$^{26}$,
M. Ahrens$^{60}$,
C. Alispach$^{32}$,
P. Allison$^{24,\: 25}$,
A. A. Alves Jr.$^{35}$,
N. M. Amin$^{50}$
% ...
\\
\\
% ...
$^{12}$ Universit{\'e} Libre de Bruxelles, Science Faculty CP230, B-1050 Brussels, Belgium \\
% ...
$^{17}$ Department of Physics, Loyola University Chicago, Chicago, IL 60660, USA \\
% ...
$^{22}$ Dept. of Physics and Astronomy, University of Canterbury, Private Bag 4800, Christchurch, New Zealand \\
% ...
$^{24}$ Dept. of Astronomy, Ohio State University, Columbus, OH 43210, USA \\
$^{25}$ Dept. of Physics and Center for Cosmology and Astro-Particle Physics, Ohio State University, Columbus, OH 43210, USA \\
$^{26}$ Niels Bohr Institute, University of Copenhagen, DK-2100 Copenhagen, Denmark \\
% ...
$^{32}$ D{\'e}partement de physique nucl{\'e}aire et corpusculaire, Universit{\'e} de Gen{\`e}ve, CH-1211 Gen{\`e}ve, Switzerland \\
% ...
$^{35}$ Karlsruhe Institute of Technology, Institute for Astroparticle Physics, D-76021 Karlsruhe, Germany  \\
% ...
$^{50}$ Bartol Research Institute and Dept. of Physics and Astronomy, University of Delaware, Newark, DE 19716, USA \\
% ...
$^{60}$ Oskar Klein Centre and Dept. of Physics, Stockholm University, SE-10691 Stockholm, Sweden \\
% ...
$^{71}$ DESY, D-15738 Zeuthen, Germany \\
% ...

\subsection*{Acknowledgements}

\noindent
USA {\textendash} U.S. National Science Foundation-Office of Polar Programs,
U.S. National Science Foundation-Physics Division,
U.S. National Science Foundation-EPSCoR,
Wisconsin Alumni Research Foundation,
Center for High Throughput Computing (CHTC) at the University of Wisconsin{\textendash}Madison,
Open Science Grid (OSG),
Extreme Science and Engineering Discovery Environment (XSEDE),
Frontera computing project at the Texas Advanced Computing Center,
U.S. Department of Energy-National Energy Research Scientific Computing Center,
Particle astrophysics research computing center at the University of Maryland,
Institute for Cyber-Enabled Research at Michigan State University,
and Astroparticle physics computational facility at Marquette University;
Belgium {\textendash} Funds for Scientific Research (FRS-FNRS and FWO),
FWO Odysseus and Big Science programmes,
and Belgian Federal Science Policy Office (Belspo);
Germany {\textendash} Bundesministerium f{\"u}r Bildung und Forschung (BMBF),
Deutsche Forschungsgemeinschaft (DFG),
Helmholtz Alliance for Astroparticle Physics (HAP),
Initiative and Networking Fund of the Helmholtz Association,
Deutsches Elektronen Synchrotron (DESY),
and High Performance Computing cluster of the RWTH Aachen;
Sweden {\textendash} Swedish Research Council,
Swedish Polar Research Secretariat,
Swedish National Infrastructure for Computing (SNIC),
and Knut and Alice Wallenberg Foundation;
Australia {\textendash} Australian Research Council;
Canada {\textendash} Natural Sciences and Engineering Research Council of Canada,
Calcul Qu{\'e}bec, Compute Ontario, Canada Foundation for Innovation, WestGrid, and Compute Canada;
Denmark {\textendash} Villum Fonden and Carlsberg Foundation;
New Zealand {\textendash} Marsden Fund;
Japan {\textendash} Japan Society for Promotion of Science (JSPS)
and Institute for Global Prominent Research (IGPR) of Chiba University;
Korea {\textendash} National Research Foundation of Korea (NRF);
Switzerland {\textendash} Swiss National Science Foundation (SNSF);
United Kingdom {\textendash} Department of Physics, University of Oxford.

\end{document}